\documentclass[10pt]{article}
\usepackage{geometry}                % See geometry.pdf to learn the layout options. There are lots.
\usepackage[parfill]{parskip}
\usepackage{amssymb}

\usepackage{amsmath}

\newcommand{\sech}{\mbox{sech}}

\title{Lax hierarchy, Solitons, Sum rules and a dual Lax Hierarchy}
\author{C. V. Sukumar \\{\em The Rudolf Peierls Centre for Theoretical Physics,}\\{\em Department of Physics, University of Oxford, Oxford OX1 3NP, U.K. }}

\begin{document}
\maketitle

\begin{abstract}
It is shown that a set of functions which characterize the Lax hierarchy of non-linear equations may be represented in terms of the eigenstates of the potential which satisfies the generalized KdV equation. Such a representation leads to sum rules relating integrals involving the soliton potential and its various derivatives to sums involving the bound state eigenvalues of the Schr{\"{o}}dinger equation for the reflectionless potential. A new hierarchy of functions, which is in a sense dual to the Lax hierarchy, is identified. It is shown that time dependent equations involving the dual functions may be established which permit solutions related to an $N$-soliton structure similar to that for the Lax hierarchy but with a different 'speed' for the solitons.

\end{abstract}

\section{Introduction}

The Korteweg-deVries (KdV) equation was first discovered in the study of water waves. The KdV and the related equations with higher order nonlinearity, which are members of the Lax hierarchy (Lax 1968), have played a fundamental role in the study of nonlinear systems because they simulate many physical systems (Scott {\it et al} 1973), admit many conservation laws and the multi-soliton solutions can be given in analytic form. The KdV equation and its generalization Kadomtsev-Petviashvli (KP) equation have also played an important role in pure mathematics because of their connection to algebraic curves, Jacobian varieties, vector bundles on curves, Schur polynomials and infinite dimensional Grassmannians (Mulase 1984). 

The connection between N-soliton solutions of the KdV equation and reflectionless potentials with $N$ bound states in non-relativistic Quantum Mechanics is well known (Kay and Moses 1956, Gardner {\it et al} 1967, Scott {\it et al} 1973). Even though the reflectionless potentials belong to a rather restrictive category, they have nevertheless proved to be useful in some areas of Physics. The property of asymptotic freedom in Quantum Chromo Dynamics guarantees that quarks can not exist freely suggesting that inter-quark potentials are confining potentials and therefore have vanishing reflection coefficients.
The zero angular momentum $s$-states of a confining potential in 3-dimension may be viewed as the odd states of a symmetric reflectionless potential. Such a picture has been used by Thacker {\it et al} (1978), Quigg {\it et al} (1980) and Quigg and Rossner (1981) to use experimental data for quarkonium systems as input information for the $N$-soliton KdV algorithm to construct quark-anti-quark potentials.

We consider higher order KdV equations given in the form (Lax 1968, Sawada and Kotera 1974, Caudrey {\it et al} 1976)
\begin{equation}
\frac{\partial U}{\partial t_m} \ + \frac{\partial L_m}{\partial x}\ =\ 0 \label{}
\end{equation}
where $[L_j]$ satisfy
\begin{equation}
L_0\ =\ U\ \ ,\ \ \frac{\partial L_j}{\partial x}\ =\ \Big(\frac{\partial^3}{\partial x^3} \ - 4 U \ \frac{\partial }{\partial x}\ -\ 2 \frac{\partial U}{\partial x}\Big)\ L_{j-1},\ \ j=1,2,..,m  \label{}
\end{equation}
and $t_m$ is the 'time' parameter of the $m^{th}$ member of the hierarchy. For example $m=1$ leads to
\begin{equation}
L_1\ = \frac{\partial^2 U}{\partial x^2} \ - \ 3\ U^2 \label{}
\end{equation}
and the third order KdV equation in the form
\begin{equation}
\frac{\partial U}{\partial t_1}\ +\ \frac{\partial^3 U}{\partial x^3}\ -\ 6 U\ \frac{\partial U}{\partial x} \ =\ 0 \ .\label{}
\end{equation}

The solutions to the non-linear equations of the KdV and other members of the Lax hierarchy may be used as potentials in linear Schr{\"{o}}dinger equations and their spectral properties may be studied by solving (using units in which mass = 1/2 and $\hbar =1$)
\begin{equation}
H\ =\ -\frac{\partial^2}{\partial x^2}\ +U \ \ ,\ H\ \psi_k \ =\ E_k\ \psi_k  \ . \label{}
\end{equation} 
The N-soliton solution of eq. (1) may be viewed as a reflectionless potential $U$ which supports $N$ bound states of the Hamiltonian operator $H$. Lax (1968) has shown that when the 'time' evolution of U is governed by eq. (1), if the 'time' evolution of the eigenstates $\psi_k$ is governed  by
\begin{equation}
B_m \ \psi_k \ =\ i\frac{\partial \psi_k}{\partial t_m} \ \ ,\ \ \psi_k(x,t)\ =\ \exp(-iB_mt)\ \psi_k(x,0)\ \label{}
\end{equation}
where the operator $B_m$ satisfies the commutator relation
\begin{equation}
[B_m,H]\ =\ -i\frac{\partial L_m}{\partial x}\ =\ i\frac{\partial U}{\partial t_m}\ =\  i\frac{\partial H}{\partial t_m}  \ ,\label{}
\end{equation}
then the eigenvalues $E_k$ of $H$ are independent of $t_m$ and the eigenstates remain normalized, but the normalization constants of the bound states and the reflection and transmission coefficients for positive energies acquire a $t_m$ dependence. $B_m$ is a Hermitian operator which may be interpreted as the generator of 'time' evolution which propagates the potential $U$ according to the KdV equation or another higher order non-linear equation arising from eqs. (1) and (2). This propagation of $U$ by $B_m$ is distinct from the usual time evolution of the Schr{\"{o}}dinger eigenstates by the Hamiltonian $H$ which propagates particles through a fixed potential. The hermiticity of $B_m$ ensures unitary 'time' evolution of the eigenstates.

For the case $m=1$, which leads to the third order KdV, the explicit form of $B_1$ is given by
\begin{equation}
B_1\ =\ i\ \Big(-4 \frac{\partial^3}{\partial x^3}\ +\ 6 U\ \frac{\partial}{\partial x} \ +\ 3 \frac{\partial U}{\partial x}\Big)\ \ . \label{}
\end{equation}
If the potential evolves in $t_1$ according to eq. (4) and the eigenstates evolve in $t_1$ according to eqs. (6) and (8) then the eigenvalues $[E_k=-\gamma_k^2]$ in eq. (5) are independent of $t_1$, but the normalization constants $[C_k]$ which determine the behaviour of $[\psi_k]$ as $x\to\pm\infty$ and the reflection coefficient $R$ for positive energies $E=k^2$  depend on $t_1$ as given by (Scott {\it et al} 1973)
\begin{equation}
C_k(\gamma_k,t_1)\ =\ C_k(\gamma_k,0)\ \exp{\big(-4\ \gamma_k^3\ t_1)}\ ,\ \ R(k,t_1)\ =\ R(k,0)\ \exp{\big(8\ i\ k^3\ t_1\big)}\ .\label{}
\end{equation}
Similar results hold for other members of the Lax hierarchy. One of the aims of this paper is to elucidate the explicit forms of $[B_m]$ and $[L_m]$ for all members of the Lax hierarchy. 

The plan of the paper is as follows : In section 2 of this paper we discuss some interesting properties of the well known $N$ soliton solutions of the Lax hierarchy using a simple self-contained approach. In section 3 we find a representation of $[L_m]$ of the Lax hierarchy in terms of the eigenstates of the N-soliton potential. We show that such a representation leads to sum rules involving the eigenvalues of the soliton potential. In section 4 we study the 'time' development of the soliton solutions and find an explicit expression for the eigenstate time evolution operator $B_m$ for the entire Lax hierarchy. The method used for deriving the time dependent equation suggests that it is possible to generalize the Lax hierarchy further. In section 5 we consider the concept of a dual to the Lax hierarchy and study the soliton structure of the first member of the new hierarchy. Section 6 contains a discussion of the main results of this paper.

In this paper the word 'time' and the symbol $t$ stand for a parameter which appears in the description of the evolution of the potential and does not refer to the coordinate $t$ which is canonically conjugate to the energy in Quantum Mechanics. With this clarification we drop the quotation marks on 'time' in the rest of the paper. 

\section{N-soliton solutions of the Lax hierarchy} 

The N-soliton solutions of the KdV equation and all other members of the Lax hierarchy can be given in many equivalent forms. At any given value of a parameter $t_m$, which is generally referred to as 'time', the $N$-soliton solution may be viewed as a reflectionless potential and the eigenstates of the Schr{\"{o}}dinger equation of the potential may be studied.  Here we consider a representation of the reflectionless potential with $N$ bound states used by Thacker {\it et al} (1978). The procedure for the construction of of reflectionless potentials with bound states is a special case of a general procedure for finding a new potential by adding bound states to a given potential (Sukumar 1986, 1987, Baye 1987) and the time evolution of the general potential will be studied in a future publication. In this paper we specialise to the case of the reflectionless potential. We suppress the dependence on $t_m$ for now and include it explicitly when studying the time evolution later in this paper. 

Even though the reflectionless potentials have been studied extensively in the literature we outline the key steps of the construction here to elucidate the connection to the generalisation of this construction in section 4. Starting from a potential $U_0=0$ with no bound states, ({\it viz}) a free particle, for which the reflection coefficient vanishes for all positive energies, the solutions to the Schr{\"{o}}dinger equation for energies $[E_k]=[-\gamma_k^2]$, given by
\begin{equation}
\lambda_k \ = \ C_k \ {\exp(-\gamma_k x)} ,\ \  k=1,2,...,N \label{}
\end{equation}
may be used to to define a matrix whose elements are %constructed from the incomplete overlap integrals of the basis of free particle solutions in the form
\begin{equation}
A_{kl} \ = \delta_{kl} \ - \ {\int_{\infty}^x \ \lambda_k(y) \ \lambda_l(y) \ dy} \ =\ \delta_{kl} \ + \ \frac{\lambda_k(x) \lambda_l(x)}{\gamma_k +\gamma_l}\ ,\ \ k,l=1,2,..,N\ . \label{}
\end{equation}

\noindent
$\bullet${\it The solutions $[\psi_l(x)]$ to the system of linear equations}
\begin{equation}
\sum_{l=1}^N {A_{kl}} \ \psi_l \ =\ \lambda_k \ ,\ \ k=1,2,....,N \label{}
\end{equation}
{\it may be constructed  and it can be shown that} by using Kramer's rule to express $\psi_l$ as the ratio of two determinants, ({\it viz}), the ratio of the determinant obtained by replacing the $l^{th}$ column of $A$ by $\lambda_k$ to the determinant of $A$. Using the property that the differential of $A_{kl}$ is $-\lambda_k \lambda_l$ and that the differential of a determinant is a sum over the N determinants constructed by differentiating each column in turn it can be shown that 
\begin{equation}
W(x)\  \equiv \ -{\sum_{l=1}^N} \lambda_l \ \psi_l  = -\sum_{l=1}^N\sum_{k=1}^N \lambda_l \big[A^{-1}\big]_{lk}\lambda_k  = \sum_{l=1}^N\sum_{k=1}^N \big[A^{-1}\big]_{lk} \frac{\partial A_{kl}}{\partial x} = \frac{\partial}{\partial x}\ {\ln {\det} A} \ .\label{}
\end{equation}

\noindent
$\bullet${\it It may be shown that $[\psi_l]$ define the N bound states of the Schr{\"{o}}dinger equation for a potential $U$ given by}
\begin{equation}
U(x) \ =\ -2\frac{\partial W}{\partial x} \ =\ -2\frac{\partial^2}{\partial x^2} \ {\ln{\det} A} \ .\label{}
\end{equation}
To prove this eq. (12) can be considered in the form
\begin{equation}
\psi_k \ =\ \lambda_k \ - \sum_{l=1}^N { \frac{\lambda_k \lambda_l}{\gamma_k + \gamma_l} \ \psi_l} \label{}
\end{equation}
and differentiated twice and rearranged to give
\begin{align}
\Big(-\frac{\partial^2}{\partial x^2} + \gamma_k^2\Big) \psi_k &=  -{\sum_{l=1}^N} \frac{\lambda_k \lambda_l}{\gamma_k+\gamma_l}\Big(-\frac{\partial^2}{\partial x^2}  +  \gamma_k^2\Big)\psi_l - \lambda_k \sum_{l=1}^N {\Big(2\lambda_l \frac{\partial \psi_l}{\partial x}-(\gamma_k+\gamma_l)\lambda_l \psi_l\Big)} \notag\\
&=-{\sum_{l=1}^N} \frac{\lambda_k \lambda_l}{\gamma_k+\gamma_l}\Big(-\frac{\partial^2}{\partial x^2}  +  \gamma_l^2\Big)\psi_l - 2\lambda_k\frac{\partial}{\partial x}\Big(\sum_{l=1}^N {\lambda_l \psi_l}\Big)\ . \label{}
\end{align}
Using eqs. (11)-(15) this can be brought to the form
\begin{equation}
A_{kl}\Big(-\frac{\partial^2}{\partial x^2} + \gamma_l^2 +U\Big) \psi_l(x) \ =\ 0 \label{}
\end{equation}
which can be used to infer that if $\det A \neq 0$ then $\psi_l$ must satisfy
\begin{equation}
\Big(-\frac{\partial^2}{\partial x^2} \ + \ U(x) \Big) \psi_l(x) \ =\ -\gamma_l^2 \ \psi_l(x) \label{}
\end{equation}
showing that $\psi_l$ is a solution in the potential $U$ for the energy $-\gamma_l^2$.

\noindent
$\bullet$ {\it Eqs. (11) and (12) may be used to show that $[\psi_l]$ satisfy the boundary conditions appropriate for bound states at $x \to \pm \infty$ and hence $[\psi_l(x)]$ are the N bound states of the potential $U(x)$ defined by eq. (14)}. In the limit  
\begin{equation}
x \to \infty\ ,\  A_{kl} \to \delta_{kl},\  {\det A} \to 1,\ \psi_k \to \lambda_k \ =\  C_k {\exp(-\gamma_kx)} \ \to 0\label{}
\end{equation}
and it may be shown that in the limit
\begin{equation}
x \to -\infty\ ,\ A_{kl} \to \frac{\lambda_j \lambda_l}{\gamma_j+\gamma_l},\  {\det A} \to \Big({\prod_{k=1}^N  \frac{\lambda_k^2}{2\gamma_k}}\Big)\  \Big(\prod_{k=1}^N \prod_{l>k}^N \Big(\frac{\gamma_k-\gamma_l}{\gamma_k+\gamma_l}\Big)^2\Big)\ \ .\label{}
\end{equation}
A similar evaluation of the determinant in the numerator of the expression for $\psi_k$ expressed as the ratio of of 2 determinants (solution to eq. (11) by the application of Kramer's rule) may be carried out to give
\begin{equation}
x \to -\infty\ ,\ \psi_k(x) \to\  \Big({\prod_{l\ne k}^N \frac{\gamma_k+\gamma_l}{\gamma_k-\gamma_l}}\Big)\  \Big(\frac{2\gamma_k}{\lambda_k(x)}\Big)\ \to\ 0 \ .\label{}
\end{equation}
The choice of the normalisation constants $C_k$ given in Thacker {\it et al} (1978)
\begin{equation}
\frac{C_k^2}{2\gamma_k}\ =\ {\prod_{l\ne k}^N \frac{(\gamma_l + \gamma_k)}{\vert(\gamma_l - \gamma_k)\vert}}\label{}
\end{equation}
corresponds to the requirement that $\vert \psi_k(R)\vert=\vert\psi_k(-R)\vert$ in the limit $R\to \infty$, which in turn corresponds to the condition that $U(x)$ be a symmetric potential. With this choice a symmetric reflectionless potential with N bound states may be constructed. It has been shown by Thacker {\it et al} that $[\psi_l]$ defined by eq. (12) are indeed normalized to unity in the interval $[-\infty,\infty]$. Here we note that since the potential defined by eq. (14) depends only on the second derivative of  the logarithm of a determinant, two determinants which defer by a factor ${\exp(ax+b)}$ will lead to the same potential whatever be the values of the constants $a$ and $b$. Hence there are many equivalent representations of the  matrix $A$ which lead to the same potential $U$ (Sukumar 1986).

\noindent
$\bullet${\it It may be shown that the reflectionless potential with $N$ bound states may be represented as a weighted sum over the $N$ bound state probability densities}. 
To establish this eq. (15) may be used to get
\begin{equation}
{\sum_{k=1}^N \gamma_k \psi_k (\psi_k - \lambda_k)} = -{\sum_{k=1}^N \sum_{j=1}^N \psi_k\lambda_k\psi_j\lambda_j\frac{\gamma_k}{\gamma_j+\gamma_k}} = -\frac{1}{2} {\sum_{k=1}^N \sum_{j=1}^N \psi_k\lambda_k\psi_j\lambda_j} = -\frac{W^2}{2} \ . \label{}
\end{equation}
Differentiation of eq. (12), multiplication from the left by $\psi_k$ and summation over $k$ yields
\begin{equation}
\sum_{k=1}^N {\sum_{l=1}^N {\psi_k\ A_{kl}\ \frac{\partial \psi_l}{\partial x}}}\ -\ \sum_{k=1}^N {\psi_k\ \lambda_k}\ \sum_{l=1}^N {\psi_l\ \lambda_l}\ =\ \sum_{k=1}^N {\psi_k\  \frac{\partial \lambda_k}{\partial x}} \ .\label{}
\end{equation}
The first term can be simplified to a sum over $\lambda_l$ multiplying the derivative of $\psi_l$ and rearranged to give
\begin{equation}
\sum_{l=1}^N {\Big(\lambda_l\ \frac{\partial \psi_l}{\partial x}\ -\ \psi_l\ \frac{\partial \lambda_l}{\partial x}\Big)}\ =\ W^2\ .\label{}
\end{equation}
The definition of $W$ in eq. (13) and $U$ in eq. (14) can then be used to show that
\begin{equation}
\sum_{l=1}^N {\Big(\lambda_l\ \frac{\partial \psi_l}{\partial x}\ +\ \psi_l\ \frac{\partial \lambda_l}{\partial x}\Big)}\ =\ -\ \frac{\partial W}{\partial x}\ =\ \frac{U}{2}\ .  \label{}
\end{equation}
Thus we can establish that
\begin{align}
\sum_{l=1}^N {\lambda_l\ \frac{\partial \psi_l}{\partial x}}\ &=\ \frac{U}{4}\ +\ \frac{W^2}{2} \ , \\
\sum_{l=1}^N {\psi_l\ \frac{\partial \lambda_l}{\partial x}}\ &=\ -\sum_{l=1}^N\gamma_l\ \psi_l\ \lambda_l\ =\ \frac{U}{4}\ -\ \frac{W^2}{2} \ . \label{}
\end{align}
Eqs. (23) and (28) may now be used to derive the expression
\begin{equation}
U\ =\ -4 \sum_{k=1}^N {\gamma_k\ \psi_k^2}\ =\ -2\frac{\partial W}{\partial x}\ =\ 2\ \frac{\partial}{\partial x} \Big(\sum_{k=1}^N \psi_k \lambda_k\Big)\ =\ -2\ \frac{\partial^2}{\partial x^2} \ln{\det{A}} \label{}
\end{equation}
exhibiting the various forms in which the potential may be expressed.

\noindent
$\bullet${\it The expression of $U$ as a weighted sum over bound state probability densities leads to a sum rule}. It is clear from the asymptotic limits given in eqs. (19) and (20) that
\begin{align}
Lt_{x\to \infty}\ W \ &=\ -Lt_{x\to \infty}\ \sum_k{C_k^2 {\exp(-2\gamma_k x)}}\ \to 0 \\
Lt_{x\to -\infty}\ W \ &=\ Lt_{x\to -\infty}\ \frac{\partial}{\partial x} {\ln{\det{A}}}\ = -2\sum_{k=1}^N {\gamma_k} \ .\label{}
\end{align}
Integration of eq. (29) then gives the sum rule
\begin{align}
\int_{-\infty}^{\infty} U(x) \ dx \ &=\ -2\ \Big(W(\infty)-W(-\infty)\Big) \ =\  -4\ \sum_{k=1}^{N} \ \gamma_k \\
&=\ -4\ \sum_{k=1}^{N} {\gamma_k \ \int_{-\infty}^{\infty} \psi_k^2 (x)\ dx} \ .\label{}
\end{align}
Eq. (32) shows that the area enclosed between the $x$-axis and a graph of the potential $U(x)$ is simply related to the binding energies of the bound states of the potential $U$. Eqs. (32) and (33) taken together confirm that the eigenstates $[\psi_k]$ defined by eq. (11) are normalized to unity in the interval $[-\infty,+\infty]$. 

Since $U(x)$ is the lowest member of the $[L_j]$ hierarchy, ({\it i.e}) $L_0=U$, it may be possible to find a representation of the higher members of the $[L_j]$ hierarchy in terms of the bound state probability densities. We examine this possibility next.

\section{Lax hierarchy and the eigenstates of U}

We now consider the differential equation satisfied by the probability density associated with the normalized bound state $\psi_k$ at energy $E_k=-\gamma_k^2$, for any potential $U$:
\begin{equation}
P_k(x) \ \equiv \ \big(\psi_k(x)\big)^2 \ ,\ \frac{\partial^2}{\partial x^2}\psi_k\ =\ \big(U\ +\ \gamma_k^2\big) \psi_k \ .\label{}
\end{equation}
It can be established that
\begin{equation}
\frac{\partial^2 P_k}{\partial x^2} \ = \ 2 U P_k \ +\ 2 \gamma_k^2 P_k \ +\ 2 \Big(\frac{\partial \psi_k}{\partial x}\Big)^2 \ .\label{}
\end{equation}
By taking a further derivative of eq. (35) a differential equation satisfied by $P_k$ can be established in the form
\begin{equation}
\Big(\frac{\partial^3}{\partial x^3} \ -\ 4 U \frac{\partial}{\partial x} \ -\ 2 \frac{\partial U}{\partial x} \Big) \ P_k \ =\ 4 \gamma_k^2  \ \frac{\partial P_k}{\partial x} \ .\label{}
\end{equation}
It may be verified that the three linearly independent solutions of this third order differential equation are $\psi_k^2, {\tilde \psi_k}^2$ and $\psi_k {\tilde \psi_k}$ where ${\tilde \psi_k}$ is the second linearly independent solution to the differential equation satisfied by $\psi_k$. The weighted sums over the probability densities defined by
\begin{equation}
Q_j \ \equiv \ -4\sum_{k=1}^N{\beta_k \big(2\gamma_k\big)^{2j} \ P_k} \label{}
\end{equation}
where $[\beta_k]$ are arbitrary functions of $k$, therefore, satisfy
\begin{equation}
\Big(\frac{\partial^3}{\partial x^3} \ -\ 4 U \frac{\partial}{\partial x} \ -\ 2 \frac{\partial U}{\partial x} \Big) \ Q_j \ =\ \frac{\partial Q_{j+1}}{\partial x} \label{}
\end{equation}
which is identical to the differential equation linking the different members of the $[L]$ hierarchy (eq. (2)). For the particular choice $[\beta_k] = [\gamma_k]$, we find using eq. (29) that
\begin{equation}
Q_0 \ = \ L_0 \ =\ U\ =\ -4 \sum_{k=1}^N{\gamma_k \ \psi_k^2} \label{}
\end{equation}
and the $[Q]$ hierarchy defined by eqs. (37) and (38) becomes identical to the $[L]$ hierarchy defined by eq. (2). 

\noindent
$\bullet${\it $[L_j]$ may be represented in terms of the bound state probability densities in the form}
\begin{equation}
L_j(x) \ =\ -2 \sum_{k=1}^N{ \big(2\gamma_k\big)^{2j+1} \ \psi_k^2(x)} \label{}
\end{equation}
{\it which leads to the sum rules}
\begin{equation}
\int_{-\infty}^{\infty}{L_j(x) \ dx} \ =\ -2 \sum_{k=1}^N{ \big(2\gamma_k\big)^{2j+1}}\ ,\ \ j=0,1,2,...\label{}
\end{equation}
The first 3 members of the $[L]$ hierarchy obtained by iteration of eq. (2) are explicitly given by
\begin{align}
L_0 \ &= \ U \\
L_1 \ &= \ \frac{\partial^2U}{\partial x^2} \ -3 U^2 \\
L_2 \ &= \ \frac{\partial^4U}{\partial x^4} \ -10 U \frac{\partial^2U}{\partial x^2} \ -\ 5\Big(\frac{\partial U}{\partial x}\Big)^2 \ +\ 10 U^3  \label{}
\end{align}
lead to the sum rules
\begin{align}
\int_{-\infty}^{\infty} U\ dx\ &= \ -4\  \sum_{k=1}^N{\gamma_k} \\
3\int_{-\infty}^{\infty} U^2 \ dx \ &= \ \ 16\  \sum_{k=1}^N{\gamma_k^3} \\
\int_{-\infty}^{\infty}\Big[5\Big(\frac{\partial U}{\partial x}\Big)^2 \ +\ 10 U^3\Big]\ dx \ &=
\ -64\ \sum_{k=1}^N{\gamma_k^5} \label{}
\end{align}
It is easy to check that these sum rules hold good for the 1-soliton potential $U=-2\gamma_1^2 {\sech}^2\gamma_1x$ for which all the integrals can be carried out analytically. 

The procedure given above may be used to establish $N$ sum rules from the first $N$ members of the $[L]$ hierarchy and if the reflectionless potential with $N$ bound states is known then the first $N$ sum rules may be used to determine the $N$ binding energies from the various integrals involving $U$ and its derivatives. This is yet another striking property of the soliton solutions of the Lax hierarchy.

\section{Time evolution of the potential and eigenstates}

\subsection{Evolution of the potential}

The potential $U$ and the bound state solutions considered in sections 2.1 and 2.2 are time independent because the basis functions in eq. (10) have no time dependence. If we consider  basis functions $\lambda_k$ which depend on a parameter $t$ in the manner given by
\begin{equation}
\frac{\partial}{\partial t}\lambda_k(x,t) \ =\ \alpha_k\ \lambda_k(x,t) \ ,\ \ \ \ 
\lambda_k(x,t) \ =\ C_k\ \exp{\Big(-\gamma_k\ x\  +\ \alpha_k\ t\ \Big)} \label{}
\end{equation}
then eqs. (11)-(18) acquire a dependence on $t$. All $\psi$ and $\lambda$ appearing in sections 2 and 3 may be viewed as corresponding to functions evaluated at $t=0$ and the time parameter can be restored by replacing all functions of $\psi(x)$ and $\lambda(x)$ by $\psi(x,t)$ and $\lambda(x,t)$. Eqs. (11) and (48) may be used to find the $t$-derivative of $A_{kl}$ in the form
\begin{equation}
\frac{\partial}{\partial t}A_{kl}\ =\ \big(\alpha_k\ +\ \alpha_l\big)\ \big(A_{kl}\ -\ \delta_{kl} \big) \label{}
\end{equation}
The $t$-evolution of $\psi_k(x,t)$ may be studied by differentiating eq. (12) with respect to $t$ and using eq. (49). These algebraic manipulations lead to
\begin{equation}
\sum_{l=1}^N A_{kl}\ \Big(\ \frac{\partial \psi_l}{\partial t}\ +\ \alpha_l\  \psi_l\ \Big)\ =\ 2\alpha_k\ \psi_k  \ .\label{}
\end{equation}
The time evolution of $\psi_l$ can now be given in the form
\begin{equation}
\Big(\frac{\partial \psi_l}{\partial t}\ +\ \alpha_l\ \psi_l\Big)\ =\ 2 \sum_{k=1}^N \big[A^{-1}\big]_{lk}\ \alpha_k\ \psi_k \label{}
\end{equation}
Multiplication of eq. (50) by $\psi_k$ and summation over $k$ yields

$$\sum_{k=1}^N \sum_{l=1}^N \psi_k A_{kl}\ \Big(\ \frac{\partial \psi_l}{\partial t}\ +\ \alpha_l \ \psi_l\ \Big) \ =\ 2\sum_{k=1}^N \alpha_k\ \psi_k^2 $$

which can be simplified using eqs. (12) and (48) to give
\begin{equation}
\sum_{l=1}^N \Big(\ \lambda_l\  \frac{\partial \psi_l}{\partial t}\  +\ \frac{\partial \lambda_l}{\partial t}\ \psi_l\ \Big)\ =\ -\frac{\partial W}{\partial t}\ =\ 2\sum_{k=1}^N \alpha_k\ \psi_k^2 \ .\label{}
\end{equation}
Using eq. (14) we can then establish the evolution equation for $U$ in the form
\begin{equation}
\frac{\partial U}{\partial t}\ =\ 4\ \frac{\partial}{\partial x}\ \sum_{k=1}^N\alpha_k\ \psi_k^2 \label{}
\end{equation}

${\bullet}$ {\it We note that this evolution equation is valid for arbitrary choices of $[\alpha_k]$ and in this method of construction of $U$ the bound state eigenvalues of $U$ are independent of $t$ which implies that this time evolution of $U$ can exist alongside a unitary evolution of the eigenstates of $U$. The potential constructed using eqs. (11)-(14) with the basis functions given by eq. (48) can now be viewed as an $N$-soliton potential. The soliton structure of $U(x,t)$ can be extracted using the procedure used by Gardner {\it et al} (1974). The non-linearity of the evolution of $U$ is implicit in eq. (53). However for special choices of $[\alpha_k]$ the non-linear character can be made explicit.} The consequences of this distinction will be further elaborated in section 5 of this paper. 

For the special choice of $[\alpha_k= 2^{2m} \gamma_k^{2m+1}]$, $m=1,2,..$, the sum in the right hand side of eq. (53) can be identified as $L_m$, a member of the Lax hierarchy defined by eq. (40). We now use the symbol $t_m$ to identify this particular $t$-evolution. Comparison with eqs. (40) and (53) then shows that the time evolution equation for $U(x,t_m)$ is now given by eq. (1). For all members of the Lax hierarchy eq. (2) may be used to express $L_m$ in terms of $U$ and its spatial derivatives and the non-linear character of the evolution equation (1) can be made explicit. The time dependent $[L_m]$ can be given in the form 
\begin{equation}
L_m(x,t_m) = -4\sum_{k=1}^N \alpha_k \psi_k^2(x,t_m) = 2\frac{\partial W}{\partial t_m}   = -2 \frac{\partial}{\partial t_m} \sum_{k=1}^N {\lambda_k(x,t_m) \psi_k(x,t_m)} = 2 \frac{\partial}{\partial t_m} \frac{\partial}{\partial x} \ln{\det{A}} \label{}
\end{equation}
which are all equivalent and provide the generalisation of eq. (29). For the case $m=0$ the evolution given by eq. (1) is simple since in this case $\alpha_k=\gamma_k$ and $L_0=U$ and $U$ and $W$ are functions of $(x-t)$.

For integer values of $m$, the potential $U(x,t_m)$ is a solution of the $m^{th}$ member of the Lax hierarchy defined by eqs. (1) and (2), which is also referred to as the $(2m+1)^{th}$ order KdV equation. We note that the limiting values of $\det A$ in the asymptotic region $x\to \pm\infty$ given by eqs. (19) and (20) are still valid when the $t_m$ dependence of $A$ is included. Hence using eqs. (19) and (20) for the asymptotic limits of $\det A$ and eq. (48) to evaluate the time derivative of ${\ln{\lambda_k}}$ it can be shown that
\begin{equation}
\int_{-\infty}^{\infty} L_m(x,t)\ dx\ =\ -2 \sum_{k=1}^N (2\gamma_k)^{2m+1} \label{}
\end{equation}
which is in agreement with eq. (41), the sum rule derived in section 3. Since the eigenstates $[\psi_k]$ in eq. (40) are normalized to unity for all $t_m$, the sum rules given by eq. (41) are independent of $t_m$ and are valid for any value of $t_m$.

\subsection{Evolution of the eigenstates}

We next examine the time evolution of the eigenstates of the time dependent potential $U(x,t_m)$ when the basis states $\lambda_k$ evolve according to eq. (48). The explicit form of the time evolution for the third order KdV is given by eqs. (6) and (8) which can be simplified using eq. (18) to the form
\begin{equation}
\frac{\partial}{\partial t_1} \psi_k(x,t_1) \ =\ -4 \gamma_k^2 \frac{\partial \psi_k}{\partial x} \ +\ \Big(2L_0 \frac{\partial \psi_k}{\partial x} \ -\ \frac{\partial L_0}{\partial x} \psi_k\Big) \ .\label{}
\end{equation}
We now show how equation (56) arises from eq. (51) for the case  $m=1$ for which $[\alpha_k=4\gamma_k^3]$ and find the form of the equation analogous to eq. (56) for other values of $m$. 

The procedure for adding $N$ bound states to a potential $U_0=0$ may be reversed so that starting from a potential with $N$ bound states in $U$ we can find a potential with no bound states in the form
\begin{align}
U_0\ &= \ U\ -\ 2\ \frac{\partial^2}{\partial x^2} {\ln{\det{B}}} \ =\ 0\\
B_{kl} \ &= \delta_{kl} \ + \ {\int_{\infty}^x \ \psi_k(y) \ \psi_l(y) \ dy} \ =\ {\int_{-\infty}^x \ \psi_k(y) \ \psi_l(y) \ dy} \label{}
\end{align}
and the two sets of functions $\lambda$ and $\psi$ are now related by
\begin{equation}
\sum_{l=1}^N{B_{kl}}\ \lambda_l\ =\ \psi_k \ \ ,\ \ \lambda_l\ =\ \sum_{k=1}^N{\big[B^{-1}\big]_{lk}\ \psi_k}  \label{}
\end{equation}
so that
\begin{equation}
-\sum_{l=1}^N \psi_l(x) \lambda_l(x) = -\sum_{l=1}^N\sum_{k=1}^N\psi_l\big[B^{-1}\big]_{lk}\psi_k = -\sum_{l=1}^N\sum_{k=1}^N\big[B^{-1}\big]_{lk} \frac{\partial B_{lk}}{\partial x} = -\frac{\partial}{\partial x} {\ln{\det{B}}} .\label{}
\end{equation}
Comparison of the two sets of equations corresponding to the addition and the removal bound states shows that the matrix $[B]$ is the inverse of the matrix $[A]$. Hence
\begin{equation}
\big[A^{-1}\big]_{lk}\ =\ B_{lk}\ =\ {\int_{-\infty}^x \ \psi_l(y) \ \psi_k(y) \ dy} \label{}
\end{equation}
which can be further simplified using the Wronskian relation
\begin{equation}
\psi_l\ \frac{\partial \psi_k}{\partial x}\ -\ \frac{\partial \psi_l}{\partial x}\ \psi_k\ =\ \big(\gamma_k^2\ -\ \gamma_l^2\big)\ {\int_{-\infty}^x \ \psi_k(y) \ \psi_l(y) \ dy} \ \ ,\ \ \  k\ne l\ .\label{}
\end{equation}
The steps taken for establishing the time evolution equation (51) starting from eq. (48) may be repeated to find the $x$ derivative of $\psi$, by the replacement $\alpha_k \rightarrow -\gamma_k$ in eq. (51), in the form
\begin{equation}
\Big(\frac{\partial \psi_l}{\partial x}\ -\ \gamma_l\ \psi_l\Big)\ =\ - 2 \sum_{k=1}^N \big[A^{-1}\big]_{lk}\ \gamma_k\ \psi_k \label{}
\end{equation}
For the KdV-Lax hierarchy $[\alpha_k]$ may be given in the form
\begin{equation}
\alpha_k\ =\ 4^m \ \gamma_k^{2m+1}\ =\ 4^m\ \gamma_k\ \big(\gamma_k^{2m}\ -\ \gamma_l^{2m}\big)\ +\ 4^m\ \gamma_k\ \gamma_l^{2m} \label{}
\end{equation}
and a factor $(\gamma_k^2\ -\ \gamma_l^2)$ may be extracted from the first term on the right hand side of eq. (64) so that the combination $\big[A^{-1}\big]_{lk}\ \alpha_k$ in eq. (51) may be expressed in terms of a Wronskian. These algebraic manipulations lead to
\begin{align}
\frac{\partial \psi_l}{\partial t}\ +\ \alpha_l\ \psi_l\ &=\ 4^m\ \gamma_l^{2m} \Big(2\sum_{k=1}^N \big[A^{-1}\big]_{lk}\ \gamma_k\ \psi_k\Big)\ +\Delta_l \\
\Delta_l\ &=\ 2^{2m+1}\ \sum_{k=1}^N \Big(\psi_l\ \frac{\partial \psi_k}{\partial x}\ -\ \frac{\partial \psi_l}{\partial x}\ \psi_k\Big)\ \Big(\sum_{j=1}^{m} \gamma_l^{2j-2}\ \gamma_k^{2m-2j}\Big) \gamma_k\ \psi_k \ .\label{}
\end{align}
Using eq. (63) for the $x$-derivative of $\psi$ and eq. (40) for $[L_k]$  it is possible to simplify the evolution equation for $\psi$ to the form
\begin{equation}
\frac{\partial \psi_l}{\partial t}\ =\ -\ \big(2 \gamma_l\big)^{2m}\ \frac{\partial \psi_l}{\partial x}\  +\ \sum_{j=1}^m \big(2\ \gamma_l\big)^{2j-2}\ \Big( -\ \frac{\partial L_{m-j}}{\partial x}\ \psi_l\ +\ 2\ L_{m-j}\ \frac{\partial \psi_l}{\partial x}\Big) \ .\label{}
\end{equation}
It is readily verified that for $m=1$ this expression reduces to eq. (56) for the third order KdV.

\noindent
$\bullet${\it The time evolution of the eigenstates of $U(x,t_m)$ for the entire KdV hierarchy is given by}
\begin{align}
\frac{\partial}{\partial t_m} \psi_k(x,t_m) \ &=\ -i\ B_{m}\ \psi_k(x,t_m) \notag\\ 
\ B_m\ \psi_k(x,t_m) \ &= -\ i\ \big(2\gamma_k\big)^{2m} \frac{\partial \psi_k}{\partial x}  +\ i\ \sum_{j=1}^{m}{\big(2\gamma_k\big)^{2j-2}  \Big(2 L_{m-j} \frac{\partial \psi_k}{\partial x}  - \frac{\partial L_{m-j}}{\partial x} \psi_k\Big)} \ .\label{}
\end{align}
We have established this equation from first principles. It is possible to verify directly that this is consistent with eqs. (5) and (7) as follows. We can use eq. (5) to show that
\begin{equation}
\big[B_m, H\big]\ \psi_k \ =\ -\big(H + \gamma_k^2\big)\ B_m\psi_k \ .\label{}
\end{equation}
The right hand side of eq. (68) consists of three types of terms and the effect of $(H+\gamma_k^2)$ acting on each of the terms can be found and simplified using eq. (5). Eq. (2) for the $[L]$ hierarchy can then be used to further simplify the expressions. These algebraic manipulations lead to the result that
\begin{equation}
\big[B_m, H\big] \ \psi_k \ =\ -i\frac{\partial L_m}{\partial x} \ \psi_k \ .\label{}
\end{equation}
We can also use eqs. (29), (40) and (68) to show that
\begin{align}
\frac{\partial U}{\partial t_m} &= -4 \frac{\partial}{\partial t_m} \sum_{k=1}^N{\gamma_k \psi_k^2} \notag\\
 &= -\frac{\partial L_m}{\partial x} + 2 \sum_{j=1}^{m}{\Big(\frac{\partial L_{j-1}}{\partial x} L_{m-j} - L_{j-1} \frac{\partial L_{m-j}}{\partial x}\Big)} \ .\label{}
\end{align}
It is easy to show that all the terms in the sum cancel each other when the sum is expanded. Thus we can verify that
\begin{equation}\big[B_m, H\big]\ \psi_k\ =\ -i\ \frac{\partial L_m}{\partial x}\ \psi_k \ =\ i\ \frac{\partial U}{\partial t_m}\ \psi_k \label{}
\end{equation}
showing that the time evolution of $\psi_k$ given by eq. (68) is consistent with eqs. (1) and (5) and (7). The time evolution operator for the eigenstates of the $N$-soliton potential of the Lax hierarchy can also be given in the form
\begin{equation}
B_m \ =\ -\ i\ \frac{\partial}{\partial x} \big(-4H\big)^m\ +\ i\  \sum_{j=1}^{m}{ \Big(2L_{m-j}\frac{\partial}{\partial x} - \frac{\partial L_{m-j}}{\partial x}\Big)\big(-4H\big)^{j-1}} \ .\label{}
\end{equation}

\section{A dual Lax hierarchy for reflectionless potentials}

We now examine the general evolution equations (48) and (53) again.  It was noted earlier that the $m^{th}$ member of the KdV hierarchy arises from the choice of $\big[\alpha_k=2^{(2m)} \gamma_k^{(2m+1)}\big]$. As emphasized earlier, eq. (53) is valid for arbitrary values of $[\alpha_k]$. This observation allows us to consider an alternate choice of $\big[\alpha_k= 2^{(-2m)} \gamma_k^{(1-2m)}\big]$ leading to a hierarchy of functions ${\bar L}_m, m=1,2,...$, which correspond to the negative values of the index $m$ of $\big[L_m\big]$ considered earlier. Thus the members of the hierarchy defined by
\begin{equation}
{\bar L}_m\ =\ -2 \sum_{j=1}^N \frac{\psi_j^2(x,t)}{(2\gamma_j)^{(2m-1)}}\ ,\ {\bar L}_0\ =\ L_0\ =\ U(x,t) \label{}
\end{equation}
satisfy
\begin{equation}
\frac{\partial {\bar L}_{m-1}}{\partial x}\ =\ \Big(\frac{\partial^3}{\partial x^3} \ - 4 U \ \frac{\partial}{\partial x}\ -\ 2 \frac{\partial U}{\partial x}\Big)\ {\bar L}_{m},\ \ m=1,2,..  \label{}
\end{equation}
and lead to the time evolution equation for the potential of the form
\begin{equation}
\frac{\partial U}{\partial t}\ =\ -\ \frac{\partial {\bar L}_m}{\partial x}\ . \label{}
\end{equation}
$\bullet${\it This hierarchy may be viewed as a dual hierarchy to the usual Lax hierarchy. In the Lax hierarchy starting from $L_0=U$ higher members are found using eq. (2) by differentiation and integration. In the hierarchy defined by eq. (75) starting from ${\bar L}_0=U$ higher members are found by solving  differential equations.} In this respect the processes for generating higher members of the sequence for the two hierarchies differ in a fundamental way. The potential $U$ is still represented by a weighted sum over the probability density of the normalized eigenstates. The method used by Gardner {\it et al} (1974) to exhibit the soliton structure of $U(x,t)$ in the context of KdV still applies and it can be shown that in the asymptotic domain, $t\to\pm\infty,\ x\to\pm\infty$, the potential becomes a superposition of $N$ localized solitons, localized around different regions of $x$, just as for the soliton solutions of the KdV hierarchy.

We now consider the $m=1$ member of this family, which can be shown to have a special significance since the non-linearity of the evolution equation for this case may be unraveled. The relevant equations for this case are:
\begin{align}
\lambda_k(x,t)\ &=\ C_k\  \exp{\big(-\gamma_k x \ +\ \frac{t}{4\gamma_k}\big)}\ \\
C_k^2\ &=\ 2\ \gamma_k \ \ {\prod_{l\ne k}^{N}\frac{(\gamma_l\ +\ \gamma_k)}{\vert(\gamma_l\ -\ \gamma_k)\vert}}\ \\
A_{kl}(x,t)\ &=\ \delta_{kl}\ +\ \frac{\lambda_k(x,t)\ \lambda_l(x,t)}{\gamma_k\ +\ \gamma_l} \\
\sum_{l=1}^N A_{kl}\ \psi_l(x,t)\ &=\ \lambda_k(x,t)\ \ ,\ \ \ k=1,2,..,N  \label{}
\end{align}
from which it is possible to construct
\begin{align}
U(x,t)\ &=\ -2\ \frac{\partial^2}{\partial x^2} {\ln{\det A(x,t)}} \ =\ -4\sum_{j=1}^{N} \gamma_j\ \psi_j^2(x,t) \\
{\bar L}_1\ &=\ -\ \sum_{j=1}^{N} \frac{\psi_j^2(x,t)}{\gamma_j} \\
\frac{\partial U}{\partial t}\ &=\ -\ \frac{\partial {\bar L}_1}{\partial x} \label{}
\end{align}
The eigenvalues $[-\gamma_j^2]$ of the Schr{\"{o}}dinger equation for the potential $U(x,t)$ defined by the above equations do not depend on time showing that the time evolution of $[\psi_j(x,t)]$ implicit in eqs. (81)-(83) is a unitary evolution. 

The evolution equation for the eigenstates can be studied by starting from eq. (51) and following steps similar to the steps taken for the Lax-KdV hierarchy (eqs. (61) - (66)). It can be shown that
\begin{equation}
\frac{\partial \psi_l}{\partial t}\ =\ -\frac{1}{4\gamma_l^2}\ \frac{\partial \psi_l}{\partial x}\ +\ \frac{1}{4\gamma_l^2}\ \Big(\frac{\partial {\bar L}_1}{\partial x}\ \psi_l\ -2\ {\bar L}_1\ \frac{\partial \psi_l}{\partial x}\Big)   \label{}
\end{equation}
which is the analogue of eq. (56) for the eigenstate evolution when $U$ satisfies the KdV equation. 

The non-linearity implicit in the evolution equation for $U$ given by eq. (83) can be made more explicit by the following reasoning. Since ${\bar L}_1$ also satisfies   
\begin{equation}
\Big(\frac{\partial^3}{\partial x^3} \ - 4 U \ \frac{\partial}{\partial x}\ -\ 2 \frac{\partial U}{\partial x}\Big)\ {\bar L}_1\ =\ \frac{\partial {\bar L}_0}{\partial x} \ =\ \frac{\partial U}{\partial x} \label{}
\end{equation}
it is evident that
\begin{equation}
\Big(\frac{\partial^3}{\partial x^3} \ - 4 U \ \frac{\partial}{\partial x}\ -\ 2 \frac{\partial U}{\partial x}\Big)\ \Big({\bar L}_1\ +\ \frac{1}{2}\Big)\ =\ 0\ .\label{}
\end{equation}
We now observe that the solutions to the Schr{\"{o}}dinger equation for the potential $U(x,t)$ at zero energy given by
\begin{equation}
\frac{\partial^2}{\partial x^2} \xi(x,t)\ =\ U(x,t) \xi(x,t) \label{}
\end{equation}
may be used to show that
\begin{equation}
\Big(\frac{\partial^3}{\partial x^3} \ - 4 U \ \frac{\partial}{\partial x}\ -\ 2 \frac{\partial U}{\partial x}\Big) \xi^2(x,t) \ =\ 0\ .\label{}
\end{equation}
Comparison of eqs. (86) and (88) shows that a solution $\xi$ to eq. (87) which satisfies suitable boundary conditions may be found such that
\begin{equation}
{\bar L}_1\ =\ \frac{(\xi^2 -1)}{2} \ .\label{}
\end{equation}
Eqs. (83) and (87) can now be combined to show that 
\begin{equation}
\frac{\partial}{\partial t}\Big(\frac{1}{\xi}\ \frac{\partial^2 \xi}{\partial x^2}\Big)\ = \ \frac{\partial U}{\partial t}\ =\ - \frac{\partial {\bar L}_1}{\partial x}\ =\ - \xi\ \frac{\partial \xi}{\partial x} \ .\label{}
\end{equation}
Thus we have shown that the nonlinear equation
\begin{equation}
\xi\ \frac{\partial^2}{\partial x^2}\ \frac{\partial \xi}{\partial t}\ -\ \frac{\partial \xi}{\partial t}\ \frac{\partial^2 \xi}{\partial x^2}\ +\ \xi^3\ \frac{\partial \xi}{\partial x}\ =\ 0 \label{}
\end{equation}
has solutions of the form
\begin{equation}
\xi^2(x,t)\ =\ 1\ -2\ \sum_{j=1}^N \ \frac{\psi_j^2(x,t)}{\gamma_j} \ .\label{}
\end{equation}

The potential $U$ is a symmetric function of $x$ at $t=0$ if $C_k$ are chosen according to eq. (78). The structure of $U, [\psi_k]$ and $\xi$ in the aymptotic domain, $t\to\pm\infty,\ x\to\pm\infty$, may be analyzed by using the same method as Gardner {\it et al} (1974) and Thacker {\it et al} (1978). In the asymptotic domain the potential $U(x,t)$ is a superposition of $N$ shifted soliton solutions of the form 
\begin{align}
u_k\ &\to\ -2\gamma_k^2\ {\sech}^2 (\gamma_k y_k\  \mp \ \delta_k)\ \\
y_k\ &= \ \Big(x\ -\ \frac{t}{4\gamma_k^2}\Big)\ ,\ \ \ \ \ \ \ \ \ \ k=1,2,..,N \\
\delta_k\ &\to\ \frac{1}{2}\ \Big(\sum_{l=1}^{k-1} \ln{\frac{\vert\gamma_l\ -\ \gamma_k\vert}{\gamma_l\ +\ \gamma_k}}\ -\ \sum_{l=k+1}^{N}\ln{\frac{\vert\gamma_k\ -\ \gamma_l\vert}{\gamma_k\ +\ \gamma_l}}\Big)  \label{}
\end{align}
localized around the $N$ different regions of $x$ where a particular $y_k$ is finite while all the other $[y_j\to\pm\infty,\ j\ne k]$. The normalized eigenstates $[\psi_k]$ are also localized around the same distinct regions of $x$ and have the form 
\begin{equation}
\psi_k\ \to \  \sqrt{\frac{\gamma_k}{2}}\  {\sech}(\gamma_k y_k \ \mp\ \delta_k) \ ,\ k=1,2,..,N \ .\label{}
\end{equation}
Hence
\begin{equation}
\xi^2\  \to\   \tanh^2(\gamma_k y_k \ \mp\ \delta_k)\ ,\ k=1,2,..,N \label{}
\end{equation}
({\it i.e}) $\xi^2$ is also localized around $N$ different regions of $x$ where $[y_k]$ are finite. Thus the non-linear equation we have derived has solutions with a structure which is related to a soliton structure similar to that for the KdV hierarchy of equations. However for  $t\to\pm\infty$ the solitons have speeds different from those of the KdV solitons because of the altered time dependence carried from eq. (77) to eq. (94).

$\xi(x,t)$ is also the solution to the Schr{\"{o}}dinger equation at energy $E=0$ for the potential given by eq. (81) which satisfies appropriate boundary conditions which are compatible with those satisfied by $[\psi_j(x,t)]$, as required by eq. (92). The zero energy solution may be directly calculated as follows. For positive energies $E = k^2$  it may be shown that eq. (15) may be extended to the form
\begin{equation}
\psi(E,x)\ =\ \exp(\pm i k x)\ -\ \sum_{l=1}^N \frac{\lambda_l(x)\ \exp(\pm i k x)}{\gamma_l\ \pm\ i k}\ \psi_l(x) \label{}
\end{equation} 
and the steps similar to the passage from eq. (15) to eq. (17) may be performed to show that the $\psi(E,x)$ so defined is indeed a solution of the Schr{\"{o}}dinger equation for energy $E$. We can now find the zero energy solution by setting $k=0$ in eq. (98) to get 
\begin{equation}
\xi(x,t)\ =\ \psi(0,x)\ =\ 1\ -\ \sum_{l=1}^N \frac{\lambda_l(x,t)\ \psi_l(x,t)}{\gamma_l}\ . \label{}
\end{equation}
By squaring this expression and using eq. (15) it may verified that eq. (99) is compatible with eq. (92). $\xi$ satisfies the boundary condition $Lt_{x\to\infty}\ \xi(x) \to 1$ and using eq. (21) it may be shown that $Lt_{x\to -\infty}\ \xi(x) \to (-)^N$. These boundary conditions are compatible with eq. (92). Thus $\xi$ may be expressed in either of the two forms given in eqs. (92) and (99). The non-linearity of the time evolution of $\xi$, as shown by eq. (91), may thus be used to characterize the implicit non-linearity in the time evolution of $U(x,t)$ through eq. (90). 

For the 1-soliton case the solution to the non-linear equation (91) is
\begin{equation}
\xi\ =\ {\tanh \big(\gamma_1\ y_1\big)}\ ,\ \ y_1\ =\ \Big(x\ -\ \frac{t}{4\gamma_1^2}\Big) \label{}
\end{equation}
and using
\begin{equation}
\phi_1\ =\ \sqrt{2\gamma_1}\ {\exp\big(- \gamma_1\ y_1\big)}\ ,\ \ \psi_1\ = \  \sqrt{\frac{\gamma_1}{2}}\ {\sech\big(\gamma_1\ y_1\big)}\ ,\ \ U_1\ =\ -2\gamma_1^2\  {\sech^2\big(\gamma_1\ y_1\big)} \label{}
\end{equation}
eqs. (86) - (92) and eq. (99) may be verified. 

For the 2-soliton case the solution to eq. (91) may be shown to be
\begin{align}
\xi\ &=\ \frac{\gamma_2\ \sinh \big(\gamma_1\ y_1\big)\ \sinh \big(\gamma_2\ y_2\big)\ -\ \gamma_1\ \cosh \big(\gamma_1\ y_1\big)\ \cosh \big(\gamma_2\ y_2\big)}{\gamma_2\ \cosh \big(\gamma_1\ y_1\big)\ \cosh \big(\gamma_2\ y_2\big)\ -\ \gamma_1\ \sinh \big(\gamma_1\ y_1\big)\ \sinh \big(\gamma_2\ y_2\big)} \\
y_j\ &=\ \Big(x\ -\ \frac{t}{4\gamma_j^2}\Big) \ \ ,\ \ j=1,2 \ .\label{}
\end{align}
and the corresponding potential and eigenstates for the case $\gamma_2 > \gamma_1$ are
\begin{align}
U_2\ &=\ -\ 2\ \frac{\big(\gamma_2^2 - \gamma_1^2\big)}{D^2} \ \Big(\gamma_2^2\ \cosh^2\big(\gamma_1\ y_1\big)\ +\ \gamma_1^2\ \sinh^2\big(\gamma_2\ y_2\big)\Big)  \\
D\ &=\ \Big(\gamma_2\ \cosh \big(\gamma_1\ y_1\big)\ \cosh \big(\gamma_2\ y_2\big)\ -\ \gamma_1\ \sinh \big(\gamma_1\ y_1\big)\ \sinh \big(\gamma_2\ y_2\big)\Big) \\
\psi_1\ &=\ {\sqrt \frac{\gamma_1}{2}}\ {\sqrt{\big(\gamma_2^2 - \gamma_1^2\big)}}\ \frac{\sinh \big(\gamma_2\ y_2\big)}{D}  \\
\psi_2\ &=\ {\sqrt \frac{\gamma_2}{2}}\ {\sqrt{\big(\gamma_2^2 - \gamma_1^2\big)}}\ \frac{\cosh \big(\gamma_1\ y_1\big)}{D}   \ \ . \label{}
\end{align}

For general $N$, $\xi$ may be constructed using eqs. (99), (80) and (77) and may be shown to be related to a Schur polynomial of order $N$.

So far we have examined the structure of the non-linearity for the $m=1$ member of the system of eqs. (74) - (76) in detail. It is clear that other members of this hierarchy also admit an $N$-soliton structure for $U$ and a unitary evolution of the eigenstates of $U$. The time evolution equation for the eigenstates of all members of the dual hierarchy can be shown to be
\begin{equation}
\frac{\partial \psi_l}{\partial t}\ =\ -\ \big(2 \gamma_l\big)^{-2m}\ \frac{\partial \psi_l}{\partial x}\ -\ \sum_{j=1}^m \big(2\ \gamma_l\big)^{-2j}\ \Big( -\ \frac{\partial {\bar L}_{m-j+1}}{\partial x}\ \psi_l\ +\ 2\ {\bar L}_{m-j+1}\ \frac{\partial \psi_l}{\partial x}\Big) \ .\label{}
\end{equation}
which is the generalization of eq. (67) for the dual hierarchy. The dual hierarchy $[{\bar L}_m]$ discussed in this paper may be viewed as complementary to the Lax hierarchy $[L_m]$.

\section{Discussion}

In this paper we have shown that for the KdV-Lax hierarchy of non-linear equations governing the time evolution of the potential $U$, a representation of $[L_m]$ in terms of the eigenstates of the potential may be found which enables the identification of the soliton structure of $[L_m]$ for $t\to\pm\infty$. Such a representation leads to certain sum rules involving all the bound states of the potential $U$. We have shown that the integral of $[L_m]$ over the entire spatial domain $[-\infty,\ +\infty]$ is proportional to the sum over the energy of the bound state raised to the power of $m+\frac{1}{2}$. With sufficient number of such sum rules it is possible to solve for the eigenvalues of the bound states entirely in terms of the integrals of $[L_m]$. We have also given explicit expressions for $[B_m]$, the time evolution operator of the eigenstates, in terms of $[L_m]$ and the derivative operator for the entire Lax hierarchy. 

It is evident from eqs. (48), (10)-(14) and (29) that, independent of the specific choice of $[\alpha_k]$, at $t=0$ there is a potential $U(x,0)$ which supports $N$ bound states of the Schr{\"{o}}dinger equation. For $t\to\pm\infty$ equations similar to eqs. (93)-(95) may be used to show that the $N$-bound state potential separates into $N$ separated packets of potentials, each with its own depth proportional to the binding energy, each supporting a single bound state or soliton for all choices of $[\alpha_k]$. For different choices of $[\alpha_k]$ the dependence of the speed of the soliton on the energy is different. For the KdV-Lax hierarchy $[\alpha_k]$ is proportional to $[\gamma_k^{2m+1}]$ giving rise to solitons which move in the time region $t\to\pm\infty$ with speeds proportional to a positive integer power $m$ of the energies of the bound states that the solitons represent. The non-linearity of the evolution equation for $U$ is of order $(2m+1)$ for this case.
 
We have identified a new hierarchy $[{\bar L}_m]$ which leads to a new type of non-linear time evolution equations for reflectionless potentials which retains the soliton structure of the Lax hierarchy and produces a unitary time evolution of the eigenstates of the Schr{\"{o}}dinger equation for the potential. An explicit expression for the time evolution operator for the eigenstates of the potential may be found for this case also in terms of $[{\bar L}_m]$ and  $[{\bar L}_m]$ may be expressed in terms of the eigenstates. For the dual Lax hierarchy identified in this paper $[\alpha_k]$ is proportional to $[\gamma_k^{1-2m}]$ giving rise to solitons with speeds proportional to a negative integer power $-m$ of the energies. The evolution equation for $U$ is now implicitly non-linear and for $m=1$ the character of the non-linearity can be unraveled in the manner discussed in section 5. For this dual hierarchy the reflectionless potentials at time $t=0$ given by $U(x,0)$ may be used to determine $\xi(x,0)$ by solving eq. (87) at $t=0$  subject to the boundary condition that $\xi^2\to1$ as $x\to\pm\infty$ and this solution may be used to find $\xi(x,t)$ by solving the non-linear evolution equation (91) and this $\xi(x,t)$ may be used back again in eq. (87) to find $U(x,t)$ and then $[\psi_k(x,t)]$. In the KdV-Lax hierarchy the potential $U$ plays the fundamental role in the sense that all other $[L_m]$ are expressed in terms of $U$ and its derivatives through eq. (2). For the dual hierarchy the zero energy solution in the potential denoted by $\xi$, satisfying the boundary conditions that $\vert\xi\vert\to1$ as $x\to\pm\infty$, plays the fundamental role in the sense that $\xi$ determines all the members $[{\bar L}_m]$ as is evident from eqs. (75) and (89). It is for $\xi$, in the $m=1$ case, that it has been possible to establish an evolution equation which is explicitly non-linear as a function of time and this is perhaps the case for all the other members of the dual hierarchy. The other members of the dual hierarchy with $m>1$ deserve further study .

There are other choices of the time evolution functions $[\alpha_k]$ which lead to other interesting evolution equations for $U$ and $[\psi_k]$, which retain the soliton structure of the KdV with the speeds of the solitons having other kinds of dependence on the energy but still providing unitary evolution of $[\psi_k]$. These aspects will be the subject of a future study.  

%\vfill\eject
\section{References}

\noindent[1] P.D.Lax, Comm. Pure. Appl. Math. {\bf 21}, 467 (1968).

\noindent[2] A.C.Scott, F.Y.E.Chu and D.W.Mclaughlin, Proc. I.E.E.E. {\bf 61}, 1443 (1973).

\noindent[3] M.Mulase, J. Diff. Geo. {\bf 19} 403 (1984).

\noindent[4] I.Kay and H.E.Moses, J. Appl. Phys. {\bf 27}, 1503 (1956).

\noindent[5] C.S.Gardner, J.M.Greene, M.D.Kruskal and R.M.Miura, Phys. Rev. Lett. {\bf 19}, 1095 (1967).

\noindent[6] H.B.Thacker, C.Quigg and J.L.Rosner, Phys. Rev. {\bf D18} 274, 287 (1978).

\noindent[7] C.Quigg, H.B.Thacker and J.L.Rosner, Phys. Rev. {\bf D20} 234 (1980).

\noindent[8] C.Quigg and J.L.Rosner, Phys. Rev. {\bf D23}, 2625 (1981).

\noindent[9] K.Sawada and T.Kotera, Prog.Theor. Phys. {\bf 51}, 1355 (1974).

\noindent[10] P.J.Caudrey, R.K.Dodd and J.D.Gibbon, Proc.Roy.Soc.Lon. {\bf A351}, 407 (1976).

\noindent[11] P.B.Abraham and H.E.Moses, Phys. Rev. {\bf A22}, 133 (1980).

\noindent[12] C.V.Sukumar, J. Phys. A: Math. Gen. {\bf 19} 2297 (1986).

\noindent[13] C.V.Sukumar, J. Phys. A: Math. Gen. {\bf 20} 2461 (1987).     

\noindent[14] D.Baye, J. Phys. A: Math. Gen. {\bf 20}, 5529 (1987).

\noindent[15] C.S.Gardner, J.M.Greene, M.D.Kruskal and R.M.Miura, Comm. Pure Appl. Math. {\bf 27}, 97 (1974).

%\noindent[16] D.Baye and J.M.Sparenberg, Phys. Rev. Lett. {\bf 73}, 2789 (1994).

%\noindent[17] P.Morse and H.Feshbach, 1953 {\it Methods of Theoretical Physics, Vol. 1}(New York: McGraw-Hill) 791-811.

%\noindent[44] A.V.Backlund, Math. Ann. {\bf 9}, 297 (1876); {\bf 19}, 387 (1882).
%\noindent[47] K.Kwong and J.L.Rosner, Prog. Theor. phys. Suppl. {\bf 86}, 366 (1986).
%\noindent[38] D.J.Korteweg and G.de Vries, Phil.Mag. {\bf 39}, 422 (1895).
%\noindent[42] C.V.Sukumar, {\it Proceedings of the Third International
%Colloquium on Differential equations,Plovdiv} (Amsterdam: VSP), 167(1993).}

\end{document}